\documentclass[12pt,english]{article}
\usepackage[latin1]{inputenc}
\usepackage{color}
\makeatletter

\providecommand{\boldsymbol}[1]{\mbox{\boldmath $#1$}}

\newcommand{\be}{\begin{equation}}
\newcommand{\ee}{\end{equation}}
\newcommand{\bea}{\begin{eqnarray}}
\newcommand{\eea}{\end{eqnarray}}
\newcommand{\ba}{\begin{array}}
\newcommand{\ea}{\end{array}}

\def\bbox{{\,\lower0.9pt\vbox{\hrule \hbox{\vrule height 0.2 cm
\hskip 0.2 cm \vrule height 0.2 cm}\hrule}\,}}
\newcommand{\dsl}{\pa \kern-0.5em /}

\def\tr{\rm tr}

\usepackage{latexsym}
\usepackage{amsmath,,calrsfs}
\usepackage{amsfonts}
\usepackage{amssymb}
\usepackage{amscd}
\usepackage{bbm}
\usepackage{fancybox}
\usepackage{cite}
\usepackage{amsmath,amsfonts,amsbsy}
\usepackage{pstricks,pst-node}
\usepackage[small,bf,hang]{caption2}
\usepackage{graphicx}
\usepackage{epsfig}
\usepackage{psfrag}
\usepackage{comment}

\font\mybb=msbm10 at 12pt
\def\bb#1{\hbox{\mybb#1}}

\font\mybb=msbm10 at 12pt
\def\bb#1{\hbox{\mybb#1}}

\usepackage{babel}
\makeatother
\begin{document}

\begin{titlepage}

\vfill

\begin{center}
\baselineskip=16pt {\Large \bf A brief history of supermembranes$^*$}
\vskip 0.3cm
{\large {\sl }}
\vskip 10.mm
{\bf Eric A. Bergshoeff$^{\dagger,1}$, Ergin Sezgin$^{\ddagger,2}$ and  Paul K.
Townsend$^{+,3}$}
\vskip 1cm
{\small
$^\dagger$
Van Swinderen Institute, University of Groningen,\\
Nijenborgh 3, 9747 AG Groningen, The Netherlands\\
}
\vspace{6pt}
{\small
  $^\ddagger$
  George P. and Cynthia W. Mitchell Institute
for Fundamental Physics and Astronomy
Texas A\&M University, College Station, TX 77843-4242, USA
}
\vspace{6pt}

{\small
  $^+$
Department of Applied Mathematics and Theoretical Physics, \\
Centre for Mathematical Sciences, University of Cambridge\\
Wilberforce Road, Cambridge, CB3 0WA, UK\\
}
\end{center}
\vfill

\par
\begin{center}
{\bf ABSTRACT}
\end{center}

``When to the sessions of sweet silent thought 

I summon up re-{\it membranes} of things past, 

I sigh the lack of many a thing I sought''. 
\bigskip

(Apologies to William Shakespeare)

\vfill

$^*$ To appear in {\sl Half a Century of Supergravity}, eds. A. Ceresole and G. Dall'Agata. 

\vfill

  \hrule width 5.cm
\vskip 2.mm
{\small
\noindent $^1$ e.a.bergshoeff@rug.nl\, , 
\noindent $^2$ sezgin@tamu.edu\, , 
\noindent $^3$ pkt10@cam.ac.uk
\\ }
\end{titlepage}



\section{Introduction} 
\setcounter{equation}{0}

In the early years of supergravity, much effort went into the construction and exploration of supergravity 
 theories in various spacetime dimensions. It was appreciated from the start that $D=11$ was the maximal dimension, 
 and the unique 11D supergravity was constructed in 1978 \cite{Cremmer:1978km}. Consideration of compactifications to 
 4D led to a revival of the Kaluza-Klein idea as a way to unify gravity with particle physics but there was no clear 
 route to a phenomenologically viable theory. Neither did it appear that supergravity theories could be ultraviolet (UV) 
 finite despite some remarkable cancellations of UV divergences. 
 
 Over the same period, advances in string theory were opening up the possibility of a role for 10D supergravity theories as low-energy effective theories for UV finite superstring theories. The initial, and still standard, formulation of superstring theory was via the Neveu-Schwarz-Ramond (NSR)  ``spinning string'' (with local worldsheet supersymmetry) combined with the Gliozzi-Scherk-Olive (GSO) projection that truncates the 
 spectrum to one with spacetime supersymmetry. For the closed IIA superstring for example (and of most relevance here) this removes 
 a tachyon and leaves the massless states of a IIA  10D graviton supermultiplet.

By the early 1980s, and especially after 1984, it was looking as though 11D supergravity enthusiasts had gone ``a dimension too far''. However,
superstring theories were defined only as weak-coupling expansions so it was still possible to imagine that there would be some non-perturbative role for 11D supergravity. An early hint of this was the 1984 construction of the IIA 
10D supergravity theory by dimensional reduction of 11D supergravity \cite{Campbell:1984zc}, but the bosonic fields of 11D supergravity are 
the 11-metric and a 3-form potential, which would couple naturally to a membrane. 

Early attempts to generalise string theory to membrane theory had not been encouraging, and it was shown in 1986 that 
the spectrum of a quantum membrane has no massless states \cite{Kikkawa:1986dm}. Supersymmetry was a potential solution to this problem but an 11D supermembrane would be needed to put this to the test and it was unclear whether such a thing could exist. No analog of the NSR formalism was available, and whereas the Nambu-Goto string equations linearise in a light-cone gauge the analogous membrane equations are intrinsically non-linear, and non-conformal.  Even if a supermembrane action could be found there would be no way to recover 
11D supergravity by imposing conformal invariance. In other words, most of the features that were crucial to the interpretation of superstring theory as a theory of gravity were unavailable to a hypothetical 11D supermembrane. 

This was the backdrop to our 1987 paper on ``Supermembranes and 11-dimensional Supergravity'' \cite{Bergshoeff:1987cm} but 
several earlier works unrelated to 11D physics were crucial to it. One was the 1984 construction of a `covariant' superstring action, now called the Green-Schwarz (GS) superstring, for spacetime dimensions $D=3,4,6,10$ \cite{Green:1983wt}. 
Another was the 1986 construction by Hughes, Liu and Polchinski of a 6D super-3-brane action as the effective action for a Nielsen-Olesen vortex solution of a 6D supersymmetric Abelian-Higgs model \cite{Hughes:1986fa}; a crucial technical contribution of this paper was a simplification of  the GS construction that allowed for $p>1$, at least in principle. We used this simplified GS construction to find an action for an 11D supermembrane in any background solution of the 11D supergravity field equations. In fact, as was soon clarified by Duff et al. \cite{Duff:1987bx},  the 
conditions on the background that we had found to be necessary for the (classical) consistency of  the 11D supermembrane action are equivalent to the 11D supergravity field equations!  We concluded with a speculation on a connection to 10D superstring theory via  dimensional reduction, and this was confirmed in the Duff et al. paper: one finds the IIA GS superstring coupled to IIA supergravity.  We also provided a general construction of super-p-branes in a D-dimensional spacetime under assumptions that limit the possible $(D,p)$ pairs, and a full classification of these possibilities, later dubbed the ``branescan'', was soon achieved \cite{Achucarro:1987nc}. 
Here we provide some details of these developments, initially focusing on `branes' in a $D$-dimensional Minkowski vacuum because the super-Poincar\'e isometries of this background provide a useful `guide-rail'. 
 
 As mentioned above, it was known that some branescan entries have an interpretation as effective actions for `extended' BPS solitons of supersymmetric 
 Minkowski-space field theories `lifted' to a higher spacetime dimension. In fact, almost all have an interpretation of this kind with
 the 10D superstrings and 11D supermembrane as the (apparent) exceptions (a `gravitational' version of this interpretation was found for them later, as we
 mention below).  This was expected from the ``fundamental'' status of 10D superstrings, and it suggested a similar status for the 11D supermembrane.  In an early review of the topic \cite{Bergshoeff:1987qx} we explored the possibility of a quantum ``supermembrane theory'' via analogies to superstring theory,  but a much better idea for the quantum theory was developed a few years later by de Wit, Nicolai and collaborators \cite{deWit:1988wri,deWit:1988xki} using special properties of membranes found and developed by Goldstone and Hoppe around five years earlier \cite{Hoppe-Thesis}. We review this here via a Hamiltonian approach to the supermembrane. 
 
Superstring perturbation theory starts from consideration of a superstring in a 10D Minkowski vacuum background. This is 
 self-consistent because the perturbation expansion in powers of the string coupling constant $g_s$ requires $g_s<<1$,  and 
 this implies that the length scale determined by the string tension (in units with $\hbar=1$) is much greater than the scale 
 set by the 10D Newton constant. The back-reaction of the string on the spacetime geometry is non-perturbative. There is no analog 
 of $g_s$ in the 11D Minkowski vacuum of 11D supergravity, and the back-reaction of a supermembrane cannot be ignored.  In particular,  
 a static planar membrane in the 11D Minkowski vacuum, can be expected to source a static membrane-like  supergravity solution with the same (super)symmetries. This solution was found by Duff and Stelle in 1990 \cite{Duff:1990xz}. The effective action for fluctuations of this BPS  `gravitational' membrane-soliton, later to be called the M2-brane,  is the 11D supermembrane action, so the distinction between membrane ``matter'' and gravity in 11D is blurred.  

Our aim here is to briefly review the evolution and development of ideas about branes in the decade 1984-1994, prior to the 
M-theory/brane revolution of 1995. We focus on the worldvolume (rather than the extreme-black-brane) perspective, 
for the sake of brevity and because it came first. An exception was the 1992 M5-brane of \cite{Gueven:1992hh} since its effective 
action was only found later; it is essentially an interacting version of 6D supersymmetric chiral 2-form electrodynamics, which puts it 
(like D-branes) outside the class of super-p-branes covered here, which have only scalar and spinor worldvolume fields. We
discuss it briefly because it is the `magnetic' counterpart to the `electric' supermembrane by a generalisation of discrete 
electric/magnetic duality.  

\section{Strings and Branes}
\setcounter{equation}{0}

We begin by setting aside any thoughts of supergravity to focus on the dynamics of relativistic branes in 
a Minkowski vacuum. The topic began with Dirac's attempt to explain the muon as an excited state of 
an electron viewed as a spherical membrane supported by electrostatic repulsion. This work introduced
the idea of a purely geometrical action proportional to the 3-volume of the  `worldvolume' $W$ swept out by the 
membrane in its time evolution. The same idea, but for a relativistic string, was introduced by both Nambu and 
Goto in the early days of string theory.  Generalising to $p$-dimensional extended objects, or $p$-branes, 
in a $D$-dimensional Minkowski spacetime, we have 
\be\label{Dirac}
I= -T \int_W d^{(p+1)}\xi \sqrt{- \det {}^*\!g}\, , \qquad {}^*\!g_{\mu\nu} = \partial_\mu X^m \partial_\nu X^n \eta_{mn} \, , 
\ee
where $\{\xi^\mu; \mu=0, 1, \dots,p\}$ are local worldvolume coordinates, and the functions  $\{X^m(\xi); m=0, 1, \dots, n\}$
specify the embedding of $W$ into a Minkowski vacuum of dimension $D=(n+1)$; its standard Minkowski metric $\eta$ (of  `mostly-plus' signature) induces the metric ${}^*\!g$ on $W$. The constant $T$ is the $p$-brane tension (as verified below). 

This Dirac-type action is invariant under diffeomorphisms of $W$. In a Monge gauge, defined by 
\be
\{X^m(\xi) \} = \{\xi^\mu, \vec X(\xi)\} \, , 
\ee
where the $(n-p)$-vector field $\vec X(\xi)$ determines local transverse displacements from a planar $p$-brane, we have 
\be\label{inducedm}
{}^*\!g_{\mu\nu} = \eta_{\mu\nu} +  \partial_\mu \vec X(\xi) \cdot  \partial_\nu \vec X(\xi)\, , 
\ee
and hence
\be\label{MongeAct}
I=  \int d^{(p+1)}\xi \left\{ -T  - \frac12 \eta^{\mu\nu} \partial_\mu \vec X \cdot \partial_\nu \vec X + \   {\rm interaction\ terms}\right\}\, .
\ee
Notice that the Monge-gauge worldvolume fields are all scalars.   There are other possibilities (the M5-brane and D-branes mentioned in the Introduction)  but they will not be discussed here, except briefly as context requires. 

The constant term in the integrand of \eqref{MongeAct}  tells us that the unperturbed planar brane has a stress-energy tensor $\mathcal{T}_{\mu\nu} = -T\eta_{\mu\nu}$, which  confirms that $T$ is both the brane's tension and its unperturbed energy density $\mathcal{E}_0$. As the speed of small-amplitude plane-wave perturbations is $\sqrt{T/\mathcal{E}_0}$, causality imposes the bound $\mathcal{E}_0\ge T$, which is saturated for a Dirac-type brane: wave disturbances travel at the speed of light. This is also evident from the term quadratic in $\partial X$ since it implies a massless wave equation for $\vec X$. Notice that there are no longitudinal waves; a corollary is that Dirac-type branes (and Nambu-Goto strings) cannot support tangential momentum. This can seen more directly from the Hamiltonian formulation, to which we now turn. 

To pass from the Lagrangian to the Hamiltonian, we first make a time-space split of the induced metric: 
let $\xi^\mu=(t,\sigma^a)$ ($a=1,\dots,p$) and write
\be
{}^*\!g_{ij} \equiv \partial_i X^m \partial_j X^n \eta_{mn}  = 
\left(\begin{array}{cc} {}^*\!g_{tt} & {}^*\!g_{ta} \\  {}^*\!g_{tb} & h_{ab}\end{array}\right). 
\ee
We assume that $W$ is foliated by $p$-dimensional ``worldspaces''  $w$ (parameterised by the time coordinate $t$); 
we write the worldspace metric as $h_{ab}$ because its inverse $h^{ab}$ must be distinguished from the space components of the inverse of the worldvolume metric ${}^*\!g_{\mu\nu}$. This distinction is needed for the identity 
\be\label{detiden}
\det {}^*\!g =  \det h \left[{}^*\!g_{tt} -h^{ab}\, {}^*\!g_{ta} {}^*\!g_{tb}\right]\, . 
\ee
Now consider the following ``phase-space'' action (restricted to closed $p$-branes):
\be\label{phasespace}
I= \int \!dt\oint_w \!d^p\sigma\left\{ \dot X^m P_m -{1\over2}\ell \left(P^mP^n \eta_{mn} + T^2 \det h\right) 
- s^a\left(\partial_a X^m P_m \right)\right\}\, , 
\ee
where $P_m$ are the components of the $D$-momentum density canonically conjugate to $X^m$,  and 
$\ell$ and $s^a$ are  Lagrange multipliers for constraints associated to time and worldspace diffeomorphism invariance; 
the latter constraint states that the momentum density tangential to $w$ is zero (as claimed above). This action can be 
shown to be equivalent to the Dirac action  by  sequential elimination of $P_m$, $\ell$ and $s^a$. 

\subsection{Light-cone gauge}

We have already discussed the Monge gauge, which breaks the  manifest `internal' spacetime Lorentz invariance of the Dirac-type action to a
worldvolume Lorentz invariance. In the current context this allows the constraints to be solved for $(P_0,P_a)$ and this yields a
phase-space action without constraints that is equivalent to an action of the form  \eqref{MongeAct}. Another useful gauge choice is the light-cone gauge.  First we define the light-cone coordinates of $D$-dimensional Minkowski spacetime by 
\be
X^\pm = \frac{1}{\sqrt{2}} \left(X^1\pm  X^0\right)\, , \qquad \bb{X} = \{ X^I; I=1, \dots, D-2\} \, .  
\ee
We then fix the time-reparametrisation invariance, and partially fix the worldspace diffeomorphism invariance, 
by choosing 
\be
X^+(t, \boldsymbol{\sigma}) = t\, , \qquad P_- =T\, . 
\ee
For this gauge choice the Hamiltonian density is $-P_+$, but the constraint imposed by $\ell$ can be solved for $P_+$. Ignoring a total time 
derivative, the phase-space Lagrangian is then found to be
\be\label{canL}
L = \oint \!d^p\sigma \left\{ D_t X^I P_I  - \mathcal{H} +   TX^- \partial_a s^a\right\} \, , 
\qquad D_t \bb{X}  := (\partial_t  - s^a\partial_a)\bb{X}\, , 
\ee
where
\be
\mathcal{H} = \frac{1}{2T}\left[|\bb{P}|^2 + T^2\det h\right] \, . 
\ee
The remaining constraint imposed by $X^-$ is $\partial_a s^a=0$, which reduces the gauge group to that of 
$p$-volume preserving diffeomorphisms; we abbreviate this to ${\rm SDiff}_p$ but it should be appreciated that 
this group depends on the $p$-brane topology. 

For $p=1$ the residual constraint implies that the (one-component) Lagrange multiplier $s$ is a function of time only. For a closed string we then have the Lagrangian
\be 
L=  \oint\!d\sigma \left\{ \dot {\bb{X} } \cdot \bb{P} - \frac{1}{2T}\left[|\bb{P}|^2 + T^2|\bb{X}'|^2\right] \right\} -
s(t) \oint\! d\sigma\,  \bb{X}' \cdot \bb{P}\, . 
\ee
The one global constraint is the level-matching condition. 

For $p=2$ the constraint imposed by $s^a$ may be solved locally in terms of a worldspace scalar field $\omega(t, \boldsymbol{\sigma})$:
\be  
s^a= \varepsilon^{ab} \partial_b\, \omega\,  . 
\ee
This is the general solution if we assume a membrane of spherical topology (otherwise, there are additional global constraints).  
Elimination of $\bb{P}$ now yields
\be
\bb{P} = D_t \bb{X} \, , \qquad  D_t \bb{X} := \dot{\bb{X}} + \{\omega, \bb{X}\}\, , 
\ee
where, for any pair of scalars $(U,V)$, 
\be
\{ U, V \} := \varepsilon^{ab} \partial_a U \partial_b V\, .    
\ee
The phase-space Lagrangian of \eqref{canL} now reduces to  
\be\label{Lag1D}
L  = \frac{T}{2}\oint d^2\sigma \left\{ |D_t \bb{X}|^2 - \sum_{I<J}\{ X^I, X^J\}^2 \right\}\, . 
\ee
This defines a 1D ``gauge-mechanics'' model with 1-form gauge potential $\omega\, dt$ and 
an ${\rm SDiff}_2$  gauge group. In fact, it is the dimensional reduction to 1D of a 
$(D-1)$-dimensional Yang-Mills theory with this gauge group.  The configurations with zero energy 
are those for which $D_t \bb{X} =0$ and $\{X^I,X^J\}=0$ for all $I,J$. These are static spherical membranes
that have collapsed to zero area.

As any scalar function on the 2-sphere can be expanded in spherical harmonics, both $\omega$ and the 
`1D fields' $\bb{X}$ can be expanded in spherical harmonics. This expansion yields the following 
infinite series of irreps of the $SU(2)$ isometry group of the round 2-sphere:
\be\label{harmonics}
[3] \oplus [5] \oplus [7] \oplus \dots \oplus [2N-1] \oplus \dots
\ee
The $SU(2)$ singlet irrep can be omitted because it corresponds to motion of the centre of mass 
of the membrane that is decoupled from the other modes. This is the starting point for
the (super)membrane quantum theory mentioned in the Introduction, which we return to later. 

We mention, in passing,  that the Dirac-type p-brane action in light-cone gauge is a 1D  gauge mechanics theory with gauge group
${\rm SDiff}_p$  but for $p>2$ it is not a gauge theory of Yang-Mills type \cite{Bergshoeff:1988hw}. There is no known way 
to view it as a limit of some sequence of 1D gauge mechanics theories with a finite-rank gauge group. In this sense, the 
$p=2$ case is special.

\section{Superstrings and Superbranes}
\setcounter{equation}{0}

 The 1984 Green-Schwarz `covariant' action for 10D superstring theories is based on a worldsheet embedded in a 10D superspace, and invariant under its super-Poincar\'e isometries. It was a generalisation of a super-Poincar\'e invariant massless superparticle action \cite{Brink:1981nb} that we return to below. For simplicity, and  maximal relevance here, we focus on the IIA superspace with coordinates $\{X^m, \theta^\alpha\}$, where $\{\theta^\alpha; \alpha=1,\dots,32\}$ are the (anticommutng) components of a  non-chiral and Majorana (`real' for real Dirac matrices) 10D spinor. The infinitesimal supersymmetry 
transformations of these coordinates with spinor parameter $\epsilon$ are\footnote{The factors of $i$ are needed because we use the 
standard convention that the product of two anticommuting `numbers' is `real'.}:
\be\label{susytrans}
\delta \theta = \epsilon \, , \qquad 
\delta X^m = -i\bar\epsilon \Gamma^m \theta\,  \qquad \left(\bar\epsilon= \epsilon^T C\right), 
\ee
where $\Gamma^m$ are the 10D Dirac matrices and $C$ is the real antisymmetric 10D charge conjugation matrix. 
A basis for super-translation invariant 1-forms on superspace is $\{d\theta^\alpha, \Pi^m\}$, where 
\be
\Pi^m = dX^m +i\bar\theta \Gamma^m d\theta \,  \qquad \left(\Rightarrow \quad d\Pi^m = i  d\bar\theta \Gamma^m d\theta\right). 
\ee 
Notice that $d\Pi^m$ is not identically zero; this is a general feature, true for any spacetime dimension $D$ because it is a 
consequence of the fact that flat superspace has a non-zero torsion. For the particular case here it follows from 
symmetry of the matrices $C\Gamma^m$ and the fact that $d\theta$ is effectively a commuting 
spinor because we are implicitly using the exterior product of differential forms. 

If  we replace $dX^m$ by $\Pi^m$ in \eqref{inducedm} we find the modified, and super-translation invariant,  induced metric
\be\label{super-g}
{}^* \! g_{\mu\nu} = \Pi_\mu ^m \Pi_\nu ^n \eta_{mn}\, , \qquad 
\left( \Pi_\mu^m = \partial_\mu X^m + i \bar\theta \Gamma^m \partial_\mu\theta\right)\, .  
\ee
Using this in \eqref{Dirac} we have a manifestly spacetime super-Poincar\'e invariant extension of the Nambu-Goto action. 
This is insufficient, however,  because the total number of fermionic variables is double that of the NSR 
superstring, and the supersymmetry transformation of $\theta$ shows that all 32 components of the 
worldsheet fermionic fields are Nambu-Goldstone fermions for 32 {\sl nonlinearly} realised spacetime symmetries. 
The spectrum will not decompose into a sum of 10D supermultiplets. 

Fortunately, there is another possible term in the action with the correct dimension; after its inclusion we have the GS superstring action
\be\label{GS}
I_{GS} = -T_1\left[\int d^2\xi \sqrt{- \det {}^*\!g} - \int_W {}^*\! b \right]\, , 
\ee
where $T_1$ is the string tension and ${}^*\! b$ is the worldsheet 2-form induced by the superspace 2-form
\be\label{b}
b=  -idX^m (\bar\theta \Gamma_m \Gamma_{11} d\theta) - \frac14 \left[\bar\theta \Gamma^m (1+ \Gamma_{11})d\theta\right]
\left[\bar\theta\Gamma_m (1- \Gamma_{11})d\theta\right]\, ,  
\ee
with $\Gamma_{11} = \Gamma^0\Gamma^1 \cdots \Gamma^9$. Notice that the bosonic truncation is still the Nambu-Goto action.
This action has a novel fermionic gauge invariance that came to be called  ``$\kappa$-symmetry''; it allows half the components of 
$\theta$ to be `gauged away', and this leads to a supersymmetric string spectrum, which coincides with that of the (GSO-projected) 
NSR superstring; at zero mass we find the 10D IIA graviton supermultiplet. 

To explain how this happens it is useful to consider the much simpler massless 10D ``superparticle'' mentioned above because its quantisation yields the massless sector of the superstring spectrum. The superparticle action can be written in the `semi-Hamiltonian' form (we continue to focus on the IIA case):
\be
I[X,\theta, P, \ell] = \int dt \left\{ \Pi_0^m P_m  - \tfrac12 \ell P^2 \right\}\, , \qquad P^2 = \eta^{mn} P_mP_n\, . 
\ee
This is manifestly supertranslation invariant and clearly describes massless particles, but it too has the problem (apparently) 
that the variables $\theta^\alpha(t)$ are Nambu-Goldstone fermions for 32 nonlinearly realised supersymmetries. However, 
this action is invariant under the following $\kappa$-symmetry' gauge transformations with 10D spacetime-spinor 
parameter $\kappa(t)$ \cite{Siegel:1983hh}: 
\be
\delta_\kappa \theta =  P\!\!\!\!\slash\,  \kappa\, , \qquad 
\delta_\kappa X^m = -i\bar\theta \Gamma^m \delta_\kappa \theta\, , \qquad \delta_\kappa\ell = -4i\bar\kappa \dot\theta\, .
\ee
Since $\det P\!\!\!\!\slash=0$ for $P^2=0$, not all components of $\kappa$ yield a non-zero $\delta_\kappa \theta$; in fact 
the total number of linearly-independent non-zero variations $\delta_\kappa \theta$ is half the number of linearly-independent 
components of $\theta$, which can then be `gauged away'. The other half are the 16 physical components of $\theta$, which are
 still Nambu-Goldstone fermions for 16 nonlinearly realized worldline supersymmetries but they are also 
superpartners to the scalar fields $\vec X$  for 16 {\sl linearly-realized}  worldline supersymmetries; in the quantum 
theory this yields a 128+128 supermultiplet, as required for the IIA 10D graviton supermultiplet.  

As stated above, the generalisation of this superparticle idea to superstrings requires the GS action \eqref{GS} but 
the 2-form $b$ is not manifestly invariant under the supersymmetry transformations of \eqref{susytrans}.
However, its supersymmetry variation is an exact 2-form as a consequence of the Dirac matrix  identity  
\be\label{Dirac-id}
(\Gamma^m\mathcal{P}_\pm)^\alpha{}_{(\beta} (C\Gamma_m\mathcal{P}_\pm)_{\gamma\delta)} \equiv 0\, , \qquad \mathcal{P}_\pm := \frac12(1\pm \Gamma_{11}). 
\ee  
It was later understood that this  ``quasi-invariance'' under supersymmetry of $b$ is a consequence of the supertranslation invariance of the superspace 3-form $h=db$ (which is therefore super-Poincar\'e invariant) and that the above Dirac matrix identity is equivalent to $dh=0$. This means that the additional term of Green and Schwarz is a kind of Wess-Zumino (WZ) term \cite{Henneaux:1984mh}. Although $h$ is exact in the de Rham cohomology of superspace it is not exact in the Chevally-Eilenberg cohomology of superspace viewed as the supertranslation group because $h=db$ is supertranslation invariant but $b$ is not \cite{DeAzcarraga:1989vh}. 

For $D\ne10$ the analogous super-Poincar\'e invariant 3-form $h$ constructed from the minimal spinor (which is Majorana-Weyl
only for $D=2$ mod $8$) will not necessarily be closed. In fact, it is closed ($dh=0$) {\sl only} for $D=3,4,6,10$ because 
a version of the identity \eqref{Dirac-id} holds in these dimensions. This allows the construction of a WZ term and hence a GS
superstring action. The same series of Dirac-matrix  identities arose earlier in a construction of super-Yang-Mills (SYM) theories with 
$2,4,8,16$ supersymmetries in their maximal spacetime dimensions $D$ \cite{Brink:1976bc}, with the same result that $D=3,4,6,10$ are the only possibilities\footnote{The result of \cite{Brink:1976bc} was actually $D=2,4,6,10$ but their 2D SYM theory is a reduction from 3D.}. 
Notice that the four allowed values of $D$ can be written as 
\be
D= 2 + {\rm dim}\, \bb{A}\, , \qquad \bb{A} = \bb{R}\, ,  \bb{C}\, ,   {\bb H}\, ,  \bb{O}\, . 
\ee
The appearance of the four normed division algebras $\bb{R},\bb{C},\bb{H},\bb{O}$ is not a coincidence, and can be understood in various ways. Of most relevance in the current context is the observation by Evans \cite{Evans:1987tm} that the Dirac matrix identities for $D=3,4,6,10$ are equivalent to the Adams trialities that both generalise the spin(8) triality required for GS-RNS equivalence and are equivalent to the normed division algebras.

\subsection{Super-p-branes}

An obvious question is whether there is a generalisation to super-p-branes of the GS-superstring action. At the time, the answer was not clear, for several reasons. One was the fact that the GS $\kappa$-symmetry parameter is (in addition  to being a spacetime spinor) a self-dual vector on the worldsheet, which has no useful $p>1$ analog.  To see why there {\sl should} be a $p>1$ generalisation it is useful to 
recall that the Nambu-Goto string provides a low-energy effective description of Nielsen-Olesen vortex-string solution 
of the 4D Abelian-Higgs model \cite{Forster:1974ga}. This model has an $\mathcal{N}=2$ supersymmetric extension, 
and in this context the effective action for the 4D vortex-string (which preserves 4 of the 8 supersymmetries) was identified by Hughes and Polchinski as the 4D $\mathcal{N}=2$ GS superstring \cite{Hughes:1986dn}. However, this model is the dimensional reduction of a 6D  $(1,0)$-supersymmetric Abelian-Higgs model, in which context the BPS vortex-string becomes a BPS vortex 3-brane that must have some GS-type 6D super-3-brane effective action. 

This reasoning led Hughes, Liu and Polchinski to look for an alternative version of $\kappa$-symmetry for the GS superstring that generalises to $p>1$, at least in principle. They found that the GS $\kappa$-symmetry transformations for the superstring 
can be rewritten in terms of a new 10D spacetime spinor parameter  $\kappa$ that is a worldsheet {\sl scalar} \cite{Hughes:1986fa}:
\be\label{ktrans}
\delta_\kappa X^m = -i \bar\theta\Gamma^m \delta_\kappa \theta\, , \qquad 
\delta_\kappa\theta = \left(\bb{I} + \Gamma\right) \kappa\, , 
\ee
where the matrix $\Gamma$ has an immediate generalisation to $p\ge1$:
\be\label{kGam}
\sqrt{-\det {}^*\! g} \, \Gamma = \frac{1}{d!}  \varepsilon^{\mu_1\cdots \mu_d}
\Pi_{\mu_1}{}^{m_1} \cdots \Pi_{\mu_d}{}^{m_d}\,  \Gamma_{m_1 \cdots m_d} \,  \qquad (d=p+1).
\ee
However, to put this to the test for $p>1$ we need a WZ term.

To construct a WZ term for $p>1$ we need a closed superspace $(p+2)$-form $h$ of the right dimension.  For a minimal spinor 
$\theta$ the only candidate super-Poincar\'e invariant $(p+2)$-form with the right dimension is 
\be\label{higher}
h=  i\Pi^{m_1} \cdots \Pi^{m_p} \left(d\bar\theta\,  \Gamma_{m_1\cdots m_p} d\theta\right). 
\ee
This is zero for some choices of $(D,p)$, depending on the properties of Dirac matrices, so we must exclude
these cases. Otherwise, we  only require $h$ to be a  closed $(p+2)$-form ($dh=0$) since this implies exactness ($h=db$) for a flat superspace 
 \cite{Evans:1988jb}.  The BPS soliton perspective suggests that $dh=0$ for $(D,p)= (5,2) \, {\rm and}\  (6,3)$, for  example. It is only necessary to verify this for the  $(D,p)= (6,3)$ case because a closed 4-form $h'$ on the minimal 5D superspace can be constructed from a closed 5-form $h$ on the minimal 6D superspace ($h'= i_\zeta h$, where $\zeta$ is the Killing vector field for translations in one of the 5 space directions).  The closure of $h$ for $(D,p)= (6,3)$ was verified in \cite{Hughes:1986fa} 
and a GS-type action for this case was shown to be $\kappa$-symmetry invariant. 

Other  ``allowed'' $(D,p)$ pairs can be guessed by consideration of other `BPS' solitons, or by other methods. For example, we used
properties of the known superfield equations for 11D supergravity to deduce the existence of a closed 4-form $h$ for D=11 and to thereby construct a $\kappa$-symmetric action for an 11D supermembrane \cite{Bergshoeff:1987cm}; an advantage of this method is that the
generalisation to include interactions with 11D supergravity is immediate, but we postpone discussion of this point until we have concluded
our super-p-brane review, which starts from the observation that $dh=0$ is equivalent to 
\be\label{p-brane-id} 
\left(d\bar\theta \Gamma^{m_1} d\theta\right) \left(d\bar\theta \Gamma_{m_1 \cdots m_p} d\theta\right) =0\, .    
\ee
At this point we need to consider some aspects of the $D$-dependence of minimal spinors and Dirac matrices. 

For  $D=4,8,9,10,11$ (mod $8$) there is a basis for the Dirac matrices for which the minimal spinor $\theta$ is real. In these cases the Dirac conjugate spinor $\bar\theta$ is also its Majorana conjugate: $\bar\theta = \theta^T C$, where $C$ is the real unitary charge-conjugation matrix. In general, this matrix has the properties
\be
C^T = \varsigma C\, , \qquad (C\Gamma^m)^T = \varepsilon C\Gamma^m\, , 
\ee
where $(\varsigma, \varepsilon)$ are two $D$-dependent signs, and 
\be
\left(C\Gamma^{(k)}\right)^T = \varepsilon^k \varsigma^{k+1} (-1)^{\frac{k(k-1)}{2}} C\Gamma^{(k)} \, , \qquad 
\Gamma^{(k)} := \{ \Gamma^{m_1\cdots m_k} \} \, .
\ee
In particular $\varepsilon=+1$ for $D=3,4,8,9,10$ (mod $8$) and the matrices $C\Gamma^m$ are symmetric. 
This guarantees that $d\bar\theta \Gamma^m d\theta$ is not identically zero (as claimed above). For $D= 5,6,7$ (mod $8$) we have 
$\varepsilon=-1$ and the matrices $C\Gamma^m$ are antisymmetric, but the minimal spinor $\theta$ is necessarily complex 
and its Dirac conjugate is not a Majorana conjugate. However, we can choose it to be a symplectic-Majorana spinor;  in the minimal case this is a complex spinor doublet of $Sp_1 \cong SU(2)$, and 
\be
\bar\theta = \theta^T \hat C\, , \qquad \hat C = C\otimes (i\sigma_2)\, .
\ee
It is now still true that $d\bar\theta\Gamma^m d\theta$ is not identically zero. We have not yet considered the chirality projection 
required for a minimal spinor in $D=6$ mod $4$, but this is easily included, as in \eqref{Dirac-id}. 

The essential point of the above paragraph is that one always can choose to define the minimal spinor in a way that 
allows us to conclude that \eqref{p-brane-id} is satisfied iff  \cite{Bergshoeff:1987cm}
\be\label{Dirac-idp}
\left(\Gamma^{m_1}\right)_{(\alpha\beta} \left(\Gamma_{m_1 \cdots m_p}\right)_{\gamma\delta)} =0\, ,  
\ee
where $\alpha$ is an index for the independent real components of a minimal spinor (whether Majorana or ``SU(2)-Majorana''), 
and where $C$ (or $\hat C$) is used to lower spinor indices (and its inverse to raise them). This Dirac-matrix condition, with symmetry on four indices because of the four $d\theta$ factors in \eqref{p-brane-id}, is apparently stronger than \eqref{Dirac-id} for $p=1$ but is in fact equivalent to it. However, the implications are slightly different for $p=1$ because then \eqref{Dirac-idp} allows the construction 
of a closed 3-form $h$ for both $\mathcal{N}=1$ (i.e. minimal) {\sl and} $\mathcal{N}=2$ superspaces, whereas only minimal supersymmetry is possible for $p>1$. 

A {\sl necessary} condition for the validity of \eqref{Dirac-idp} for any $p\ge1$  can be found by contracting the left-hand side 
with $(\Gamma^n)^{\alpha\beta}$. For $p>1$ this yields 
\be\label{bscan}
2(D-p-1) = \frac12 \tr(\bb{I})\, . 
\ee
where $\bb{I}$ is the identity matrix acting on minimal spinors.  This relation imposes a severe limitation on the possible $(D,p)$ values because 
the right-hand side grows exponentially with $D$. In fact, the possibilities (including $p=1$) are \cite{Achucarro:1987nc}
\be\label{branescan}
\begin{aligned}
\bb{R}:& \quad (D,p) = (3,1)\qquad (4,2) \\
\bb{C}:& \quad (D,p) = (4,1) \qquad (5,2) \qquad (6,3) \\
\bb{H}:& \quad (D,p) = (6,1) \qquad (7,2) \qquad (8,3) \qquad (9,4) \qquad (10,5) \\
\bb{O}:& \quad (D,p) = (10,1) \quad\  (11,2) \nonumber
\end{aligned}
\ee
This classification is known as the ``branescan''.  Each of the four $\bb{R},\bb{C},\bb{H},\bb{O}$ GS-superstrings 
is the beginning of a series of $p>1$ cases. We have derived this from an implication of \eqref{Dirac-idp}, so a verification 
that \eqref{Dirac-idp} holds is still required; this task was completed in \cite{Achucarro:1987nc}. 
As explained above, the $\bb{C}$ series have an interpretation as effective actions for a BPS vortex. 
The $\bb{R}$ and $\bb{H}$ series have a similar interpretation (e.g. a kink-membrane solution of a 4D Wess-Zumino model \cite{Achucarro:1988qb} or an instanton-fivebrane of a 10D SYM model \cite{Zizzi:1983zz,Townsend:1987yy}). 

There is a simple interpretation of the branescan that is already implicit in our earlier discussion of the Monge-gauge.  
The bosonic worldvolume fields in the Monge gauge are the $(D-p-1)$ scalar fields $\vec X$ and the number of fermionic 
worldvolume fields is $\frac12 \tr(\bb{I})$, in the notation of \eqref{bscan}, where the factor of $\tfrac12$ is due to 
$\kappa$-symmetry. However, the fermions obey first-order equations and the bosons  second order equation, so \eqref{bscan} implies that the the phase-space dimensions of the bosons and fermions are the same. This is the fermi-bose matching that is required of a supersymmetric field theory, and all supersymmetric free-field  field theories in dimensions $d\ge3$ for which all bosons are scalar fields are easily classified: we seek all  values of $d=p+1$ with $D-p-1$ scalar fields. For example, for $d=3$ (and hence $p=2$) we can have $1,2,4,8$ scalar fields and hence $D=4,5,7,11$ as the possible spacetime dimensions for the supermembrane.  From our discussion of the light-cone gauge for membranes, at the conclusion of the previous section,  we might expect to find that a $D=4,5,7,11$ supermembrane in a suitable generalisation of the light-cone gauge is a $D=3,4,6,10$ SYM with ${\rm SDiff}_2$ gauge group dimensionally reduced to 1D. This is true \cite{deWit:1988wri}, and we thus have a membrane interpretation of the above-mentioned result concerning the spacetime dimensions possible for SYM theories. 

\subsection{Topological charges in the supersymmetry algebra}

We conclude this section with the resolution of an apparent clash between the spacetime supersymmetry algebra and 
the half-preservation of  supersymmetry by super-p-branes. The standard supersymmetry algebra does not allow any static soliton solution to preserve any number of supersymmetries; what makes it possible for BPS solitons is the appearance in the supersymmetry algebra of a topological central charge; e.g. magnetic charge for a 4D SYM-Higgs theory, or an instanton charge for 5D SYM theory \cite{Witten:1978mh}. In a higher dimension these become antisymmetric tensor charges that are central only with respect to the supertranslation algebra. In general, for a BPS p-brane-soliton we get  a p-form central charge in the supertranslation algebra with 
a definite coefficient (up to a sign), and this result must be duplicated in the supersymmetry algebra of the corresponding 
 effective super-p-brane action.  We now explain how this happens. 
 
For each branescan entry we have a GS-type action of the form 
\be\label{GSp}
I  = -T\!\int \!d^p\xi \sqrt{- \det {}^*\!g} +  I_{WZ} \, , 
\ee
where $T$ is the $p$-brane tension and $I_{WZ}$ is the WZ term constructed from the closed super-Poincar\'e invariant $(p+2)$-form $h$.
By construction, a super-p-brane in  a $D$-dimensional Minkowski vacuum background is invariant under the super-Poinca\'e isometries of 
this background. We have already explained how the WZ term allows a static planar p-brane to preserve half of the supersymmetries, 
and this means that it must also modify the supersymmetry algebra. It does so precisely because its supersymmetry variation is not 
zero but a total derivative. Using the notation and conventions sketched above, one finds that \cite{deAzcarraga:1989mza}
\be\label{susyalg}
\{ Q_\alpha, Q_\beta\} = 2(\Gamma)_{\alpha\beta}^m P_m + 2 (\Gamma^{m_1\cdots m_p})_{\alpha\beta} Z_{m_1\cdots m_p}\, , 
\ee
where the p-form charge $Z$ is (the exterior product of forms is implicit)  
\be
Z^{m_1\cdots m_p} = T\int_w {}^*(dX^{m_1}  \cdots dX^{m_p})\, . 
\ee
This is a  ``topological''  charge. It is zero if $w$ is a closed surface that is deformable to a point. It is also 
the charge associated to the following {\sl identically}  `conserved' worldvolume current:
\be
J^{\mu m_1\cdots m_p} :=  T\varepsilon^{\mu\nu_1\cdots \nu_p} \partial_{\nu_1}X^{m_1} \cdots \partial_{\nu_p} X^{m_p} 
\qquad (\partial_\mu J^{\mu m_1\cdots m_p} \equiv 0)\, . 
\ee

To illustrate how this resolves the puzzle, we consider a static planar static membrane, such that 
$X^m= (t, \boldsymbol{\sigma}, \vec 0)$. For this case the only non-zero components of $P_m$ and $Z_{mn}$ are 
\be
P^0  = T \int_w \!d^2\sigma \, , \qquad Z_{12} = T \int_w \!d^2\sigma\, .  
\ee
We then find (choosing $C= \Gamma^0$) that \eqref{susyalg} reduces to 
\be\label{M}
\{ Q_\alpha, Q_\beta\} = 2T M_{\alpha\beta}\left(\int_w\! d^2\sigma\right)\, , \qquad M_{\alpha\beta} := \left[ \bb{I} - \Gamma_{012}\right]_{\alpha\beta} \, .  
\ee
Of course, the factor $\int_w d^2\sigma$ is infinite for an infinite planar brane, but we may periodically identify to make it finite. 
With this understood, we see that the number of zero eigenvalues of the matrix $M$ is the number 
of non-zero spinors $\epsilon$ satisfying the condition
\be
\Gamma_{012} \epsilon = \epsilon\, . 
\ee
As  $\Gamma_{012}$ is traceless and squares to the identity, this number is $32/2 =16$; i.e. the static planar membrane preserves
half the supersymmetry of the IIA supergravity vacuum. This result (which agrees with our earlier conclusions) depends crucially on the coefficient for the WZ term. For any other choice the matrix $M$ becomes one with 32 positive eigenvalues (implying no linearly realised  supersymmetries) or worse: one with 16 negative eigenvalues, implying negative energy configurations (and a non-unitary quantum theory).

\section{The 11D supermembrane}\label{sec:gravity}
\setcounter{equation}{0}

We now focus on special features of the 11D supermembrane, starting with its surprising relation 
to the 11D supergravity field equations. Then we explain how the light-cone gauge leads to a 
{second-quantised} supermembrane theory. Finally, we discuss the gravitational back-reaction 
and the M2-brane supergravity solution, with the promised mention of the M5-brane.  
 
\subsection{Relation to 11D supergravity}

Let us return to the  11D supermembrane of the branescan  and replace the flat superspace with an 11D superspace appropriate for the superspace description of 11D supergravity.  The first step is to introduce frame 
one-forms $\{ E^A; A = a, \alpha\}$, where $a=0,1, \dots,9$ and $\alpha =1, \dots, 32$. For 
arbitrary superspace coordinates $Z^M$,  their components define the supervielbein $E_M{}^A$. The leading component of 
the superfield $E_m{}^a$ is the spacetime vierbein, from which we can construct the spacetime metric $g_{mn}$. 
No metric on superspace is defined, or needed; the geometry of superspace (in the standard version for which the tangent-space structure group is the Lorentz group) is determined  not by the Riemann tensor but by the torsion 2-forms $T^A= dE^A$; for {\sl flat} (Minkowski) 
superspace the only non-zero component is 
\be
T_{\alpha\beta}{}^a {\big |}_{\rm flat}  = \left(\Gamma^a\right)_{\alpha\beta}  \qquad 
\Leftrightarrow \quad E^A {\big |}_{\rm flat}  = (d\theta^\alpha, \Pi^a)\, . 
\ee
An induced metric on the worldvolume can be defined by 
\be\label{gen-ind}
{}^*\! g_{\mu\nu} = {}^*\! E_\mu {}^a {}^*\! E_\mu {}^b \eta_{ab}\, , \qquad  {}^*\! E_\mu {}^a = \partial_\mu Z^M E_M{}^a\, . 
\ee
For flat superspace this metric is the one of \eqref{super-g}, so we now have a Dirac-type term that 
includes a coupling of the 11D supermembrane to the spacetime metric $g$. 

We also need a WZ term, which must now be constructed from a closed {\sl superspace} 4-form $F=dA$ of the form
\be
F =  E^aE^b E^\alpha E^\beta  (\Gamma_{ab})_{\alpha\beta}\, . 
\ee
This 4-form reduces to $h$ of \eqref{higher} for flat superspace and was known to be a closed 4-form for a general
solution of the 11D superspace equations for 11D supergravity field equations \cite{Brink:1980az}. This allows us to 
write down an 11D supermembrane action of the form 
\be\label{pact}
I = - T \int_W {\rm vol}({}^*\!g)   + T\int_W {}^*\! A\, ,  
\ee
where ${}^*\! A$ is the worldvolume 3-form potential induced by $A$. The $\kappa$-symmetry transformations 
that reduce to those of \eqref{ktrans} for flat superspace are 
\be
\delta_\kappa E^\alpha = [(1+\Gamma)\kappa]^\alpha\, , \qquad \delta_\kappa E^a =0\, . 
\ee
The matrix $\Gamma$, which is traceless and squares to the identity, is now given by 
\be
\sqrt{-\det {}^*\! g} \, \Gamma = \frac{1}{6}  \varepsilon^{\mu\nu\rho}
E_\mu{}^{a} E_\nu{}^b E_\rho{}^c\,  \Gamma_{abc} \, . 
\ee

However, the above construction only guarantees $\kappa$-symmetry in the flat superspace limit.  
Beyond this, the allowable superspace backgrounds must be determined by {\sl imposing} $\kappa$-symmetry. 
This leads to a set of constraints on the components of the torsion 2-forms 
$T^A$ and the 4-form $F$. These include 
\be
T_{\alpha\beta}^a = (\Gamma^a)_{\alpha\beta}\, , \qquad F_{\alpha\beta ab} = - \frac16 (\Gamma_{ab})_{\alpha\beta}\, , 
\ee
which are expected because they confirm that the action reduces for flat superspace to the (11,2) case of the $(D,p)$ 
branescan. The higher-dimension constraints required for $\kappa$-symmetry turn out to be the superspace field 
equations of 11D supergravity \cite{Bergshoeff:1987cm,Duff:1987bx}.  The classical dynamics 
of an 11D supermembrane is consistent only for backgrounds that solve the 11D supergravity field equations! 

\subsection{Light-cone gauge quantisation}

We have already seen how the $D=4,5,7,11$ supermembrane in light-cone gauge is the 1D gauge-mechanics model  found by 
dimensional reduction of the $D=3,4,6,10$ SYM theory for gauge group ${\rm SDiff}_2$. We have also seen that, for a membrane of spherical topology, the 1D fields $\bb{X}(t)$ are functions on the 2-sphere that can be expanded in spherical harmonics; i.e. the 
infinite series of $SU(2)$ irreps of \eqref{harmonics}.  Compare this expansion with the decomposition of the adjoint irrep of 
$SU(N)$ into a sum of irreps of its principal $SU(2)$ subgroup:
\be 
 [3] \oplus [5]\oplus \cdots \oplus [2N-1]\, . 
\ee
This is a truncation of the infinite series of $SU(2)$ irreps of \eqref{harmonics}. Obviously,  the Lie-bracket relations between the irreps of the finite sum, which collectively define the Lie algebra of $SU(N)$, cannot coincide with the  Lie-bracket relations between the irreps in the infinite sum of \eqref{harmonics}. However, these relations among first $n<<N$ terms differ by terms of order $1/N$, which allows us to think of $SU(N)$ for large $N$ as an approximation to ${\rm SDiff}_2$ for the sphere. In this approximation, $X^I$ and 
$\omega$ become $SU(N)$ matrices, and 
\be 
\frac{T}{2}\oint \! d^2\sigma \to  \tr \, , \qquad \{U,V\} \to -i\,  [\, , ] \, . 
\ee
For example, for $N=2$, the functions $X^I$and $\omega$  become 3-vectors and the Lagrangian of \eqref{Lag1D} 
reduces to 
\be\label{SU2Lag}
L= \frac12 \left\{\sum_I |\dot {\bf X}^I + \boldsymbol{\omega} \times {\bf X}^I| - \sum_{I<J} |{\bf X}^I \times {\bf X}^J|^2\right\}\, . 
\ee
Notice that the potential is zero when all 3-vectors ${\bf X}^I$ are co-linear. This is an infinite ``valley''  with sides that become increasingly steep away from the origin. In the quantum theory the zero point fluctuations lead to a confining potential, implying the absence of zero energy states. 
However, we have not yet introduced supersymmetry. 

For a supermembrane we get an SYM theory and the above mentioned zero point energies cancel. This implies that the spectrum of the supermembrane has zero energy states but also that it is continuous from zero \cite{deWit:1988xki}. This was initially viewed as a supermembrane instability 
that put an end to the idea of a quantum 11D supermembrane theory \cite{deWit:1989yb}. The reason for the continuous spectrum is that one may deform a membrane to one with arbitrarily long `spikes' of arbitrarily small total area and hence arbitrarily low energy.
This is obviously true classically but supersymmetry justifies this intuition in the quantum theory. This means that there is no real distinction between one membrane and many membranes because all membranes can be connected by `tubes' of arbitrarily low energy. This suggests a re-interpretation of supermebrane quantum theory as an intrinsically second-quantized multi-membrane theory. 

This re-interpretation could have been proposed around 1990, but it had to wait for the rediscovery of the light-cone 11D supermembrane 
as a Matrix model for D0-branes in IIA superstring theory; this came much later and is therefore beyond the scope of this article. However, we conclude 
with a mention of one other feature that confirms the idea. The Lagrangian of \eqref{SU2Lag} can be understood as 
describing the centre of mass for two particles, but they are exchanged by the $Z_2$ central subgroup of $SU(2)$.
Since this $SU(2)$ is a gauge invariance, this exchange has no physical effect and the particles are therefore identical, 
and this argument generalises to $SU(N)$.

\subsection{The M2-brane and the M5-brane}

We stated earlier that the $\bb{R},\bb{C},\bb{H}$ series of branescan branes can be viewed as low-energy effective actions for 
BPS solitonic branes of Minkowski-space field theories, and we explained how this can be understood from the 
fact that a particle-like soliton in a field theory obtained by dimensional reduction on $T^p$ becomes a p-brane
of the unreduced theory. The $\bb{O}$ series has a similar interpretation but now we need to consider `gravitational solitons', 
{\it alias} extreme (zero Hawking temperature) black holes, and their higher-dimensional counterparts: extreme black branes. We should expect a planar static 11D supermembrane to be the source for a solution of 11D supergravity that is asymptotically Minkowski in directions transverse to  the plane and have the same (super)symmetries. This solution\footnote{Of the {\sl bosonic} equations, but setting all fermions to zero is always self-consistent.} (in fact, a multi-membrane generalisation of it) was found in 1990 by Duff and Stelle \cite{Duff:1990xz}:
\be\label{M2}
ds^2 = H^{-\frac23} ds^2 (\bb{E}^{1,2}) + H^{\frac13} ds^2(\bb{E}^8) \, , \qquad F= {\rm vol}(\bb{E}^{1,2}) \wedge dH^{-1}\, , 
\ee
where $H$ is a harmonic function on $\bb{E}^8$. The one-membrane solution in polar coordinates for $\bb{E}^8$ corresponds to the 
choice $H= 1+ q/r^6$ (for membrane charge $q$). The singularity at $r=0$ was initially viewed as the membrane source but it was later shown to be a coordinate singularity at the event horizon of its maximal analytic extension, although there is still a membrane-type singularity in the interior spacetime  \cite{Duff:1994fg}. Moreover,  dimensional reduction to 10D yields the  ``extreme'' string solution of 10D IIA supergravity, found earlier in 1990 by Dabholkar et al. \cite{Dabholkar:1990yf}; the fact that its strong-coupling singularity becomes the non-singular M2-brane event horizon 
 in 11D was an early hint of a non-perturbative role for the 11D supermembrane \cite{Duff:1994fg}. 

The M2-brane, as the 11D supergravity membrane solution came to be called, can be viewed as a  ``gravitational' BPS solitonic-membrane.  
It is BPS not only because it preserves 16 of the 32 supersymmetries of the Minkowski vacuum but also because it saturates a 
lower bound on the tension/charge ratio; remarkably, an `FFA' term in the action with precisely the coefficient in the 11D supergravity 
action is essential for the existence of this bound \cite{Gibbons:1994vm}.  It also breaks the Poincar\'e invariance of the 11D 
vacuum to a product of  a 3D Lorentz invariance with $SO(8)$. Together, these properties imply that its low-energy effective description 
is as an 11D supermembrane. 

Generically, the field equations for a $(p+1)$-form potential $A$ with a $(p+2)$-form field-strength $F=dA$ 
can be rewritten (by an exchange of field equations with Bianchi identities) as `dual' field equations for a $(\tilde p +1)$-form
potential $\tilde A$ with a  $(\tilde p+2)$-form field-strength $\tilde F=d\tilde A$, where
\be\label{emdual}
\tilde p = D-p-4\, . 
\ee
This is essentially a generalisation of the $\bb{Z}_2$ electromagnetic duality of electromagnetism, but instead 
of electric and magnetic `poles' we now have `electric' p-branes and `magnetic' $\tilde p$-branes.  An application to 
11D supergravity might suggest that the `electric' M2-brane should have a `magnetic' dual M5-brane. However, 
the presence of the `FFA' term in the 11D supergravity action discouraged this idea because it implies that 
the 3-form potential $A$ cannot be replaced by a 6-form, {\sl even in the field equations}.  Nevertheless, a fivebrane 
solution preserving 16 supersymmetries was found by G\"uven  in 1992 \cite{Gueven:1992hh}:
\be\label{M5}
ds^2 = H^{-\frac13} ds^2 (\bb{E}^{1,5}) + H^{\frac23} ds^2(\bb{E}^5) \, , \qquad F= \star dH\, .  
\ee
where $H$ is a harmonic function on $\bb{E}^5$ and $\star$ indicates a Hodge-dual on  $\bb{E}^5$. It was clear that
the effective worldvolume action for this M5-brane solution must reduce in a Monge gauge to some 6D 
chiral field theory for the $(1,0)$ supermultiplet with five scalars $\vec X$ and a 2-form potential with a 
self-dual 3-form field-strength \cite{Gibbons:1993sv}, but it took time for this to be found. This takes us 
beyond our remit but in order to explain what the M5-brane `naysayers' overlooked, we remark that the 
11D supergravity field equations allow a 6-form potential $\tilde A$ to be defined even though these equations 
still involve the 3-form $A$, and the 11D super-fivebrane couples to {\sl both} $A$ and $\tilde A$. 

Yet another reason to interpret both the M2 and M5 brane solutions of 11D-supergravity  as BPS solitons is that they
 both interpolate between maximally supersymmetric `vacua', one being the 11D Minkowski spacetime and the other being 
either the  $AdS_4\times T^7$ vacuum  (for M2)  or the $AdS_7\times S^4$ vacuum (for M5) \cite{Gibbons:1993sv}. Both solutions (we 
pass over here the very similar $AdS_5\times S_5$ IIB supergravity vacuum solution associated to the D3-brane)
can be viewed as extremal black branes with an event horizon that  coincides, in a near-horizon limit, with the Killing 
horizon of the AdS space in `horospherical' coordinates. The `brane' itself (which should really be viewed as a multi-brane 
since it is being treated as a classical extended object) separates an internal KK region 
from an external Minkowski vacuum. The KK spectrum always includes a `singleton' supermultiplet, which `lives' only at the AdS boundary, 
so we should expect these to be the centre-or mass degrees of freedom of the M2 or M5-brane. 
This is in fact true, e.g. \cite{Nicolai:1984gb}, and this supports the earlier ``membrane at the end of the universe'' 
idea \cite{Bergshoeff:1987dh} that was a precursor to the AdS/CFT correspondence.

However, there are still some reasons to view the supermembrane as `fundamental'  and  its `magnetic' dual  superfivebrane 
as `solitonic. As mentioned above, there is a membrane-type singularity behind the M2-brane event horizon; the maximal analytic extension 
is similar to that of the extreme Reissner-Nordstrom black hole solution of the 4D Einstein field equations. In contrast, the maximal analytic 
extension of the M5-brane solution is completely regular, and remains so when one considers the multi-fivebrane solution in which the 
horizon geometries are perturbed by the presence of other fivebranes \cite{Gibbons:1994vm}.

\section{Further Reflections}

Following the construction of the minimal 4D supergravity in 1976, the most immediate challenges were
the coupling to generic matter supermultiplets and the construction of supergravity theories with more supersymmetries. 
A large part of the latter challenge was equivalent to the challenge of constructing supergravity theories in higher 
dimensions. The highest dimension was $D=11$ and the unique  11D supergravity theory found in 1978 became a 
natural candidate for a new unified field theory. A Kaluza-Klein (KK) revival was soon underway. 

In all this initial excitement an apparently minor novelty, that attracted only sporadic attention, was the appearance of $(p+1)$-form gauge 
potentials for various values of $p$.  It was well-established (by Kemmer in the 1930s and, in much more detail,  by Ogievetsky
and Polubarinov in the 1960s) that a massless spin-zero particle (in 4D spacetime) could be described by a 2-form gauge theory,  
but there was no clear advantage over the scalar field description. In higher dimensions, however,  the notion of `spin' must be 
extended and gauge theories of $(p+1)$-form potentials with $p>0$ cannot generally be replaced by scalar, vector or symmetric-tensor fields. 
In some supergravity theories they have Chern-Simons-type couplings but the only minimal couplings occur for
 1-form potentials in  ``gauged'' supergravity theories. There is a good reason for this: a $(p+1)$-form gauge potential couples 
 `minimally' to a $p$-brane, and $p$-brane `charges' for $p>0$ are not carried by the local fields of supergravity. 
 
 In the early days of string theory, Kalb and Ramond had observed that the 2-form gauge potential in the massless string spectrum 
 couples minimally to the string itself \cite{Kalb:1974yc}. This made it natural to suppose that there should exist a supermembrane to 
 which the 3-form gauge potential of 11D supergravity can be minimally coupled. However, this idea was in tension with the idea of 11D 
 supergravity as a stand-alone unified theory,  and also with string-theory (in part because of the important role of 2D conformal invariance that does not 
 generalise to membranes). It is therefore an amusing irony that  the resurrection of the idea of an 11D supermembrane, and its construction 
 in our paper of 1987, was largely due to an advance in superstring theory: the 1984 GS-superstring alternative to the standard NSR 
 formulation. 
 
 By the early 1990s a revival of interest in supergravity was underway. Supergravity string-like solutions that are sources 
of the Kalb-Ramond 2-form potential were found in 1990 \cite{Dabholkar:1990yf}.  This was followed, in 1991, by black-p-brane solutions that 
are sources for the Ramond-Ramond (RR) $(p+1)$-form  potentials of 10D N=2 supergravity theories \cite{Horowitz:1991cd}, but as there are no 
RR p-branes with $p>1$, at least in string perturbation theory, the significance of these solutions was unclear.  Supergravity was 
still not viewed as any kind of  guide to the quantum superstring theory.  Any feature of supergravity that was in conflict with expectations from 
string theory was generally viewed as an artefact of a low-energy approximation. An early exception to this attitude was the idea, 
with support from supergravity,  that the conjectured S-duality of super-Yang-Mills theories should also apply to the Calabi-Yau-compactified 
heterotic string theory \cite{Font:1990gx}. This was taken up again in 1993 by Schwarz and Sen for $T^6$ compactifications \cite{Schwarz:1993mg}; in this context the S-duality conjecture is more compelling because of restrictions imposed in 4D by $\mathcal{N}=4$ supersymmetry. 

In the context of $T^6$-compactified  Type-2 10D superstring theories, the T-duality and conjectured S-duality are
both subgroups of a discrete version of the $E_{7,7}$ Cremmer-Julia duality group of 4D $\mathcal{N}=8$ supergravity \cite{Cremmer:1979up}, so a natural conjecture is that this discrete $E_7(\bb{Z})$ group is a duality group of the superstring theory \cite{Hull:1994ys}. One striking implication of this (unifying) U-duality
conjecture was that it requires ``wrapping'' modes of the RR p-branes to be part of the massive `BPS' spectrum in addition to KK modes and 
string winding modes. Their combined charges are in the ${\bf 56}$ of $E_7(\bb{Z})$, which decomposes under the $Sl(2,Z)\times SO(6,6)$ product of the S-duality and T-duality groups as $({\bf 2},{\bf 12}) \oplus ({\bf 1},{\bf 32})$, which puts the (16+16) electric and magnetic charges of the RR-branes in a singlet of the S-duality group, thereby confirming their non-perturbative status.  Moreover, if these modes are present then they are also present in the 
$T^7$-compactified spectrum of any putative 11D quantum gravity theory with 11D supergravity as its low-energy effective theory,
which can be construed both as a self-consistency check on the idea of such a theory, and evidence that it must include a supermembrane. 

We leave the story here. Much of the M-theory sequel is well-known, and the 11D supermembrane plays a part in this and many other
subsequent topics, about which we hope to report on some time before the 50th anniversary of the supermembrane. For the 
moment, we are pleased to have been able to contribute to the commemoration of 50 years of supergravity.

\subsection*{Acknowledgments}

The work of E.S. is supported in part by the NSF grant PHYS-2413006. The work of E.B. is supported by the Croatian Science Foundation 
project IP-2022-10-5980, {\sl Non-relativistic supergravity and applications}.

\end{document}